KX
\documentstyle[12pt,epsf]{article}
\def\ll{\label}
\def\re{\ref}
\def\c{\cite}
\def\b{\begin}

\def\r1{(\ref{$1})}
\def\ot{\otimes}

\def\ti{\tilde}

\def\ba{\begin{array}{c}}
\def\e{\end}
\def\sk{\smallskip}
\def\ea{\end{array}}
\def\pr{\prod}
\def\ni{\noindent}
\def\si{\sigma}
\def\da{\dagger}
\def\De{\Delta}
\def\de{\delta}
\def\bet{\beta}
\def\ov{\over}
\def\ha{{1\over 2}}

\def\l{\left}
\def\l({\left(}
\def\r){\right)}
\def\r{\right}
\def\rw{\rightarrow}
\def\om{\omega}
\def\la{\lambda}
\def\al{\alpha}

\def\be{\begin{equation}}
\def\bc{\begin{center}}
\def\ec{\end{center}}
\def\bit{\begin{itemize}}
\def\eit{\end{itemize}}
\def\ee{\end{equation}}
\def\ed{\end{document}}
\def\bea{\begin{eqnarray}}
\def\eea{\end{eqnarray}}

\def\bfl{\begin{flushleft}}
\def\efl{\end{flushleft}}
\def\bfr{\begin{flushright}}
\def\efr{\end{flushright}}

\begin{document}
\bibliographystyle{unsrt}

\title{ Quantum Integrable Systems:
Basic Concepts and Brief Overview} 

\author{
Anjan Kundu \\  
  Saha Institute of Nuclear Physics,  
 Theory Group \\
 1/AF Bidhan Nagar, Calcutta 700 064, India.
 }
\maketitle
%

\begin{abstract} 
An overview of the  quantum integrable systems (QIS) is presented. 
Basic concepts  of the theory are highlighted
stressing on the unifying algebraic properties, which not only
helps to  generate systematically the representative  Lax operators of different  
models,
 but also solves the related eigenvalue problem in an almost model
independent way.
Difference between the approaches in the 
 integrable
ultralocal and nonultralocal quantum  models are explained
and the interrelation between the QIS and other subjects are   
 focussed on.

\end{abstract}


\section{Introduction}
\setcounter{equation}{0}

The theory and applications of nonlinear integrable systems  
is a vast subject with wide range of applications in diverse fields
including biology, oceanography, atmospheric science, optics,  plasma etc.
etc. The quantum aspect of the subject is a relatively new development.
However the theory of quantum integrable systems (QIS) today has grown up
into an enormously rich area with fascinating relations with variety of
seemingly unrelated disciplines.
The QIS
in one hand  is intimately connected with abstract mathematical
objects like noncocommutative Hopf algebra, braided algebra, universal
${\cal R}$-matrix etc. and on the other hand  is related to the concrete
physical models in low dimensions including quantum spin chains, Hubberd
model, Calogero-Sutherland  model as well as  QFT models like sine-Gordon (SG),
nonlinear Schr\"odinger equation (NLS) etc.  The deep linkage with the 
 stat-mech problems, conformal field theory (CFT), knots and braids etc.
is also a subject of immense importance.

In  giving the account of this whole picture within this
short span of time, I am really faced with 
the problem of Tristam Shendi
\c{brussell}, who 
in the attempt of writing his autobiography 
needed two years for describing  the rich experience of
the first two days of his life and thus left us  imagining  when he would 
acomplish his mission. 
 Therefore I  will limit myself only to certain aspects of this 
 important 
  field and will be happy if  it can arouse  some of your  interests 
 in this fascinating subject.

We have to start     possibly  from  an  August day in   1934,
when 
a British engineer--historian,  John Scott Russell had a chance
encounter with  a strange
stable  wave in the Union  canal of Edinburgh \c{russel}.
Such {\em paradoxically} stable solutions will be observed again 
after many many years 
in the famous computer experiment of Fermi, Ulam and Pasta \c{fup}.
However only   in the  mid-sixties
such  fascinating phenomena
will be understood  fully as the solutions  of nonlinear 
  integrable systems 
and  named  as {\it Solitons} \c{krus65}.

Formulation of the  integrable theory of  quantum  systems 
  started only in late seventies \c{fadrev}, though today many research
groups all over the globe are engaged in  active research in this field.

Mathematical basis of classical integrable systems was laid down
mainly through the
works of Sofia Kawalewskaya, Fuchs, Painlev\'e, Liuoville and others
\c{lakh93}.
There are many  definitions of integrability ; 
 we however
  adopt the notion of  integrability
in the Liuoville sense, where integrability means the existence of
{\it action-angle} variables. That is,  if in a Hamiltonian system
$\ H [p(x,t), q(x,t)] \ $ given by the nonlinear equation
\be \dot {p}= - {\delta H  \over \delta q}, \quad
\dot {q}=  {\delta H  \over \delta p}, \ll{nle}\ee
it is possible to find a canonical transformation $\quad \left(p(x,t),
q(x,t)\right) \rightarrow \left( a(\la), b(\la,t)\right), \quad$
such that the new Hamiltonian  becomes dependent only on the action
variables, i.e.  $\ H= H[a(\la)] \ $,
then the  system may be called {\em completely integrable}.
In this case the dynamical
equations:
$ \quad \dot {a}= - {\delta H  \over \delta b} =0, \quad
\dot {b}=  {\delta H  \over \delta a} = \om , \quad  $
can be trivially solved and moreover we  get $a(\la)$ as the generator
of  the  conserved quantities. 
 The number of such independent set of conserved quantities 
in integrable systems  coincides with the degree
of freedom   of the system and   in  field models
it  becomes 
infinite.  One of these conserved quantities 
 may  be considered as the Hamiltonian. 
The inverse scattering method (ISM) \c{soliton} is an   effective 
method for solving nonlinear equations,
The important feature of ISM is that, instead of attacking the
nonlinear equation (\re{nle}) directly, it constructs 
the corresponding linear
scattering problem
\be
{\cal T}_x(x,\lambda)= L(q(x,t),p(x,t), \lambda)~ {\cal T}(x,\lambda)
,\ll{laxeq} \ee
where  the Lax operator $L(q,p, \lambda)~$ depending on  the fields $q, p$  
and the {\it spectral parameter} $\la$ contains all  information
about the original nonlinear system and may serve therefore 
as the representative of a concrete model. The field $q$ in ISM
acts as the scattering potential.
The 
aim of ISM   is to  find presizely the canonical mapping from 
 the action-angle variables to  the original field and  using it
to construct  the exact solutions for the
 original nonlinear equation. Soliton 
 is a special solution, which corresponds
 to the reflectionless ($b(\lambda )
=0$) potential.

\section{Examples of  integrable systems}
\setcounter{equation}{0}

Let us see  some concrete examples
of the Lax operators associated with well known models to get an idea about
the structure of this immensely important object 
in the  integrable systems.

\noindent I. {\it Trigonometric Class}:\smallskip

\ni
1. Sine-Gordon (SG) model (Equation and Lax operator)  

\begin{equation}
u(x,t)_{tt}- u(x,t)_{xx} = \frac {m^2}{\eta} \sin(\eta u(x,t)), \quad
  {\cal L}_{SG}  = \left( \begin{array}{c} ip , \qquad
  m  \sin (\la-\eta u) \\
   m  \sin (\la+\eta u),  \qquad -ip
    \end{array} \right), ~~p={\dot u}
\ll{sg}\end{equation}
                     \noindent
2. Liouville model (LM)    (Equation and Lax operator)
\be
u(x,t)_{tt}- u(x,t)_{xx} = \frac {1}{2} e^{2\eta u(x,t)}, \qquad
  {\cal L}_{LM}  = i\left( \begin{array}{c} p , \qquad
   \xi e^{\eta u} \\
  \frac {1}{\xi}e^{\eta u},  \qquad -p
    \end{array} \right).
\ll{lm}\end{equation}

\noindent
3. Anisotropic $XXZ$ spin chain  (Hamiltonian and Lax operator)   
\bea
{\cal H} &=& \sum_n^N(\si_n^1 \si_{n+1}^1+\si_n^2 \si_{n+1}^2 +\cos \eta
\si_n^3 \si_{n+1}^3), \nonumber \\
{L_n }(\xi) &=&
  \left( \begin{array}{c}
  \sin(\la + \eta  \si_{n}^3) ,
 \qquad  2i\sin \al \si_n^- \\
 2i\sin \eta \si_n^+    ,\qquad
  \sin(\la - \eta  \si_{n}^3) ,
    \end{array} \right)\ll{XXZ}\eea

\noindent II. {\it Rational Class}:
\sk

\ni
  1. Nonlinear Schr\"odinger equation (NLS)  (Hamiltonian and Lax operator)

\begin{equation}
i\psi(x,t)_{t}+ \psi(x,t)_{xx} + \eta (\psi^\da(x.t)\psi(x,t))\psi(x,t)=0,
      \quad
{\cal L}_{NLS}(\la)  = \left( \begin{array}{c} \la ,
\quad \eta^\ha \psi \\
\eta^\ha  \psi^\da, \quad -\la
    \end{array} \right).
\ll{nls}\end{equation}

\ni
2. Toda chain   (TC)   (Hamiltonian and Lax operator)
\be
H=\sum_i\left(\ha p^2_i
+ e^{(q_i-q_{i+1})}
\right),\qquad
L_n(\la) = \left( \begin{array}{c}
  p_n
-\la \qquad    e^{q_n}
 \\- e^{-q_n
}
 \qquad \quad 0
          \end{array}   \right).
\ll{rtoda}\end {equation}

Let us note  the following important points on the 
structure of the above Lax operators.
\begin{description}

\item  {i)} 
The Lax operator description   generalises also to the quantum case
\c {fadrev,qism}.
 Its elements   depend, apart from the spectral
parameter $\la$, also on the field operators $u,p$ or $\psi,  \psi^\da$ etc
and therefore the quantum $L(\la)$-operaors 
 are unusual matrices with noncommuting matrix elements. This intriguing
feature leads to nontrivial underlying algebraic structures in QIS.

\item {ii)} The off-diagonal elements (as $\psi, \psi^\da$ in (\re{nls}) and 
$\si^-,\si^+$ in (\re{XXZ})) involve {\em creation} and 
{\em annihilation}  operators while the diagonal terms are the number like 
operators. It is obvious that   under matrix multiplication also this property
is maintained, which  has important implications, as we will see  below.

\item {iii)} The first three models, though diverse looking, belong to the 
same trigonometric class. Similarly   the rest of the models
 represents  
the rational class. This fact signals about 
 a fascinating universal behaviour  in integrable systems based on its rich
algebraic structure.

\end{description}

\section{Notion of  quantum integrability }
\setcounter{equation}{0}

Note that the Lax operators are defined locally at a point $x$, or if we
discretise the space, at every lattice point $i$. However,
since the integrability is related to the conserved quantities, which are
indeed global objects, we also have
 to define some global entries out of the local
description of the Lax operators.  
Such an object can be formed by matrix 
multiplying Lax operators at all points as
\be
~~T(\la)=\prod_{i=1}^N L_i(\la)~=
 \left( \begin{array}{c}
  A(\la)
 \qquad    B(\la)
 \\ C(\la)
 \qquad D(\la)
          \end{array}   \right).
\ll{monod}\end {equation}
Here the global operators $B(\la), C(\la)$ are related to the angle
like variables, while $ 
A(\la), D(\la)  $ are like action variables and   
$~\tau(\la)=tr T(\la)~=A(\la)+ D(\la)$
generates the conserved operators : $\ln \tau (\la)=\sum_j C_j
\la^j$. For ensuring integrability one must show for the conserved
quantities that
 $~ [ H, C_m]=0,   [ C_n, C_m]=0 ,$ which is achieved by a key requirement
on the Lax operators (for a large class of models) given by the 
matrix relation known as the Quantum Yang--Baxter equation (QYBE)
\be~~ R_{12}(\lambda , \mu)~  L_{1i} (\lambda)~  L_{2i}(\mu )
~ = ~  L_{2i}(\mu )~ L_{1i }(\lambda)~~ R_{12}(\lambda , \mu),
\ll{qybel}\ee~
with the appearance of a $4 \times 4$-matrix $R(\la,\mu)$ with c-number functions of
spectral parameters, satisfying in turn the YBE
\be~~ R_{12}(\lambda , \mu)~  R_{13} (\lambda, \gamma)~  R_{23}(\mu ,\gamma)
~ = ~  R_{23}(\mu ,\gamma)~ R_{13}(\lambda,\gamma)~~ R_{12}(\lambda , \mu).
\ll{ybe}\ee~
Due to some deep algebraic property related to the Hopf algebra 
the same QYBE  also holds  globally:
\be~~~ R_{12}(\lambda , \mu)~ { T_1} (\lambda)~ { T_2}(\mu )
~ = ~  { T_2}(\mu )~ { T_1 }(\lambda)~~ R_{12}(\lambda , \mu),\ll{qybeg}\ee
with the notations $~T_1=T \ot I,~ T_2 =I \ot T. ~~$
Taking the trace of  relation (\ref{qybeg}), ( since under the trace
$R$-matrices can rotate cyclically and thus cancel out)  one gets $ [
\tau(\lambda),\tau(\mu)]=0 ,$ establishing 
the commutativity of $C_n$ for different $n$'s and hence proving
 the quantum integrability.

The QYBE (\ref{qybeg}) represents in the matrix form  a set of 
 commutation relations between  {\it action} and {\it angle} variables,
which can be obtained by inserting in (\re{qybeg}) 
matrix  (\re{monod}) for $T$ and
the    solution for quantum  $R(\la,\mu)$-matrix, which may be given by 
\begin {equation}
R(\lambda) = \left( \begin{array}{c}
f(\lambda) \ \qquad \ \qquad \ \qquad \\
    \quad \ 1 \  \ \ \ f_1 \ \quad  \\
     \qquad \ f_1 \ \ \ \ \ \
     1 \ \quad \\
        \qquad \ \qquad \ \qquad \ f(\lambda)
          \end{array}   \right).
\ll{R-mat}\end {equation}
The solutions  are usually of only two different types
(we shall not speak here of more general elliptic
 solutions), {\it trigonometric} 
with
\be
 f={ \sin(\la+\eta) \ov   \sin \la },
~~~f_1  ={ \sin \eta \ov   \sin \la}  
\ll{trm}\ee
and the {\it rational}  with  \be
f = { \la+\eta \ov  \la},~~ 
 f_1 = {\eta \ov \la}  
\ll{rrm}\ee

\section{Exact solution of eigenvalue problem through algebraic Bethe ansatz}
\setcounter{equation}{0}

Such generalised commutation relations dictated by the QYBE are of the form
\bea
A(\la) B(\mu)&=&  f(\mu-\la)  B(\mu) A(\la) + \cdots, \ll{ab} \\
D(\la) B(\mu)&=&  f(\la-\mu) B(\mu) D(\la) + \cdots, \ll{db} 
\eea
together with the trivial commutations for  $ [A(\la), A(\mu)]  = 
[B(\la), B(\mu)]  = [D(\la), D(\mu)]  = [A(\la), D(\mu)]  = 0$ etc.

It is now important to note that the $off$-diagonal element $B(\la)$ acts
like an creation operator (induced by the local creation operators of 
 $L(\la)$ as argued above). Therefore if one can  solve
 the quantum eigenvalue
problem
\be
  H\mid m>= 
  E_m\mid m>
\ll{energy}\ee 
or more generally
\be
  \tau(\la)\mid m>= 
  \Lambda_m(\la)\mid m>
\ll{evp}\ee
the eigenvalue problem for all  $C_n$'s can be obtained 
simultaneously by simply  expanding
$\Lambda(\la)$ as
\be
  C_1 \mid m>= 
  \Lambda_m'(0)\Lambda_m^{-1}(0)\mid m>
,~~C_2 \mid m>= 
  (\Lambda_m'(0)\Lambda_m^{-1}(0))'\mid m>
\ll{evpcn}\ee
etc.
The $m$-particle state $\mid m>$ may be considered to be created 
by $B(\la_i)$ acting $m$ times on the pseudovacuum $\mid 0>$:
 \be \mid m>= B(\la_1) B(\la_2)\ \cdots B(\la_m)\mid 0>
\ll{phim}\ee
Therefore for solving (\ref{evp}) through the Bethe ansatz we have to drag $
\tau(\la)=A(\la)+ D(\la)$ through the string of $B_{\la_i}$'s 
without spoiling their structures (and thereby preserving the eigenvector)
and hit finally the 
pseudovacuum giving $A(\la)\mid 0>=\al(\la)\mid 0>$ and $
D(\la)\mid 0>=
\bet(\la)\mid 0>$. Notice that for this purpose (\ref{ab},\ref{db})
coming from the QYBE  are
the right kind of relations. (the other type of unwanted
  terms  are usually
present in the LHS in lattice models ((as $\cdots$ in
(\ref{ab},\ref{db})), which however may be removed
by the  Bethe equations for determining the parameters $\la_j$, induced by
the periodic boundary condition.
 In case of
field models such terms are absent and  $\la_j$ become arbitrary.)
As a result we finally solve the eigenvalue problem to yield
\be \Lambda_m (\la) =
 \pr_{j=1}^m f(\la_j-\la)
\al(\la)+
\pr_{j=1}^m f(\la-\la_j)
\bet(\la).
\ll{lambda}\ee

\section{Universality in integrable systems}
\setcounter{equation}{0}
The structure of the eigenvalue $\Lambda_m (\la)$  reveals the curious fact
that apart from the $\alpha(\al),\bet(\al)$ factors 
it depends basically on the nature of the function $f(\la-\la_j)$, which
 are known trigonometric or rational functions 
 given by (\re{trm}) or (\re{rrm}) and thus is the same for 
all  models belonging to the same class. Model dependence is reflected
only in the form of $\al(\la)$ and $ \bet (\la)$ factors.
 Therefore the models like SG, Liouville
and $XXZ$ chain belonging to the trigonometric class share  similar  type of
eigenvalue relations (with specific  forms for  $\al(\la)$ and $ \bet (\la)$). 
This  deep rooted
 universality feature in integrable systems carries important
  consequences. 
\subsection{Generation of models}
One may start with the trigonometric solution (\re{trm}) for
the $R$-matrix and consider a generalised model with Lax operator

\be L_t(\la)= \left( \b{array}{c}\sin (\la+ \eta s^3),
\qquad \sin \eta S^- \ \\
\sin \eta S^+ , \qquad \sin  (\la -\eta s^3) \e{array}
\right) \ll{lslq2}\ee
 with the abstract operators $
 s^3,S^{\pm} $ belonging to the quantum algebra (QA) $U_q(su(2))$:
\begin {equation}
 [s^3,S^{\pm}] = \pm S^{\pm} ,\quad
  [ S^+, S^-]= [2 s^3]_q.
\ll{sl2qa}\end {equation}
where $  [x]_q= \frac {q^x-q^{-x}}{q -q^{-1}}
=  \frac {\sin( \alpha x)}{\sin \alpha} , \ \ q = e^{i \alpha}. $
Following the above Bethe ansatz procedure the eigenvalue would 
naturally be like (\re{lambda}) and different realisations of the quantum
algebra (\re{sl2qa}) would derive easily the eigenvalues for concrete models
belonging to this class. At the same time  the Lax
operators of these models can also be  generated 
from (\re{lslq2}) in a systematic way. 

For example, 
\be
S^\pm=\ha \si ^\pm, s^3=\ha \si ^3
\ll{qasc}\ee
constructs from (\re{lslq2}) the Lax operator of the spin-$\ha$ XXZ-chain
and describes the Bethe-ansatz solution for the suitable choice of 
 $\al(\la)$ and $ \bet (\la)$.
Similarly,
\be
 s^3_n=u_n, \  S_n^-= g(u_n) e^{i \Delta p_n}, S^+_n=(S^-_n)^\dag
,\ll{qalsg}\ee
with $\ g(u_n) = [1+ \ha m^2 \Delta ^2 \cos 2 \eta (u_n+\ha)]^\ha \ $
yields (lattice) sine-Gordon 
model. At $\De \rw 0$ one gets the SG field model with the 
Lax operator obtained as $L_n=I+\De {\cal L} (x) +O(\De) $.

All the
conserved quantities of the model including the Hamiltonian can in principle
be derived using the Lax operator.  
In fact a more general form of the ancestor Lax operator than that of
(\re{lslq2})
exists  corresponding to the same trigonomrtric $R$-matrix,the explicit form
of which can be found in ref. \cite{construct}. Concrete  realisations of such ancestor
models generates various quantum integrable models (in addition to  those
already mentioned) like quantum Derivative NLS, Ablowitz-Ladik model,
relativistic Toda chain etc. The Bethe ansatz solutions for these models
also can be obtained (with specific case-dependent 
difficulties) following
the scheme for their ancestor model, which as mentioned above is almost
model independent and  same for all  models of the same class.

At $q \rightarrow 1$ limit , $R_{trig} \rightarrow R_{rat} $ and given by the
elments (\re{rrm}). The ancestor model also reduces to the corresponding
rational form
\be L_r(\la)= \left( \b{array}{c} \la+ \eta s^3,
\qquad  \eta s^- \ \\
 \eta s^+ , \qquad \la -\eta s^3 \e{array}
\right). \ll{lsl2}\ee
  The underlying QA (\re{sl2qa}) becomes the standard $su(2)$
algebra
\begin {equation}
 [s^3,s^{\pm}] = \pm s^{\pm} ,\quad
  [ s^+, s^-]= 2 s^3.
\ll{sl2a}\end {equation}
 Such rational ancestor model (or with more generalised form
\cite{construct})
in its turn reduces also to quantum
integrable models like spin-$\ha$ XXX chain, NLS  model, Toda chain etc.
For example,
 spin-$\ha$ representation  $s^a= \ha \sigma^a$ gives the Lax operator of 
XXX chain from (\re{lsl2}), while
the  mapping from spin to bosonic operators given by 
Holstein-Primakov transformation
\be 
s^3= s- \De \psi^\da \psi, s^-=    \De^\ha(2 s- \De\psi^\da
\psi)^\ha\psi^\da, \ \ s^+= (s^-)^\da
\ll{hpt}\end {equation}
leads  to the quantum integrable  Lattice NLS model.
Similarly the Toda chain model can also be derived from the ancestor dodel
\c{construct}. 
The Bethe ansatz solutions for these desendant models
 also mimics the scheme for their
ancestor model with rational $R$-matrix. 

Thus for both the trigonometric and rational classes 
        one can construct the Lax operators and solve the eigenvalue problem
exactly through Bethe ansatz in a systematic way. This unifies diverse
models of the same class as decendants from the same ancestor model 
and at the same time realisations like
(\re{qalsg}) gives a criterion for defining {\it integrable nonlinearity} as
different nonlinear realisations of the underlying QA.
This fact of the close relationship between seemingly diverse models 
also explains in a way the {\it strange} statements often met
in other contexts like  'Quantum Liuoville model is equivalent to 
spin $ (- \ha)$ anisotropic  chain' \c{liuFad} or
'High energy scattering of hadrons in QCD is described by the Heisenberg
model with noncompact group' \c{lipatov}.

\subsection {Algebraic structure of integrable systems}

The underlying QA , as mentioned before, exhibits Hopf algebra
property. The most prominent characteristic of it is the coproduct structure
given by 
\be
\De (s^3)= s^3 \otimes I+I\otimes s^3, \ \ \
\De (S^\pm) =S^\pm \otimes q^{s^3}+ q^{-s^3}\otimes S^\pm, 
\ll{copr}\ee
This means that  if $S^\pm_1=S^\pm \otimes I$  and 
 $S^\pm_2= I \otimes S^\pm$ satisfy the QA separately, then their tensor
prodct   $\De (S^\pm)$ given by (\re{copr}) also satisfies the same algebra.
This Hopf algebraic property of the QA induces the crucial transition  from the
local QYBE (\re{qybel}) to its tensor product given by the global equation 
(\re{qybeg}), which in turn guarantees the quantum integrability of the
system as shown above.                   
 
The  QIS  described above 
are known as the  {\it ultralocal} models. 
They  are the 
standard and the most studied ones.
 The ultralocality is refered to their common property that the Lax
operators of all such models at different lattice points $i \not = j$
commute: $ [L_{1i},L_{2j}]=0.$ Note that this is consistent with the 
property: $ [ S^a_{i},S^b_{j}]=0$ for the generators of the  quantum algebra 
described above. This  ultralocality is actively used for transition from
the 
local to the global QYBE, i.e.   
in establishing  their quantum integrability.

Note that 
 the standard matrix multiplication rule 
   \begin {equation}
({A} \otimes  {B})(C \otimes D) = (AC \otimes BD)
\ll{mult}
\end {equation}
which
 holds
due to the commutativity of $B_2=I \otimes B$ and $C_1=C \otimes I,$
   remains also  valid
for the ultralocal Lax operators
 with the choice
 \be A=L_{i+1}(\la),
 B=L_{i+1}(\mu),
 C=L_{i}(\la),
 D=L_{i}(\mu) . 
\ll{mull}\ee
Therefore starting from the local QYBE (\ref{qybel}) at $i+1$ point, 
  multiplying with the same
relation at $i$ and subsequently using (\ref{mult})  with (\ref{mull})
 one globalises
the QYBE and repeating the step for $N$ times  obtains finally the
global QYBE (\ref{qybeg}).
This in turn leads to 
the commuting 
traces $\tau(\la)=Tr T(\la)$ giving  commuting 
 conserved quantities $C_n, n=1,2,\ldots,N.$

\section { Nonultralocal
models and braided extension of QYBE}
\setcounter{equation}{0}

However, There exists another class of models, known as 
 nonultralocal models (NM) with 
  the 
property $[L_{1i}, L_{2j}] \not = 0,$
for which
the trivial multiplication property (\ref{mult}) of quantum algebra
fails and it  needs
generalisation to the braided algebra \cite {majid},  
where the noncommutativity of $B_2, C_1$ could be 
 taken into account. Consequently
 the QYBE  should be generalised for such models.
 Though many celebrated models, e.g.
quantum KdV model, Supersymmetric models, nonlinear $\sigma$ models, WZWN
etc. belong to this class, apart from  few     
   \cite{maillet,kunhla} not enough studies have been devoted to this problem.
The  generalised  QYBE for nonultralocal systems
with the inclusion of  braiding matrices 
  $Z$ (nearest neighbour braiding)   and $\tilde Z$ (nonnearest neighbour
   braiding ) may be given by
\begin{equation}
{R}_{12}(u-v)Z_{21}^{-1}(u,v)L_{1j}(u)\tilde Z_{21}(u,v)L_{2j}(v)
= Z_{12}^{-1}(v.u)L_{2j}(v) \tilde Z_{12}(v,u)L_{1j}(u){R}_{12}(u-v).
\ll{bqybel}\end{equation}
In addition, this must be  complemented 
 by the  braiding relations  
\begin{equation}
 L_{2 j+1}(v)Z_{21}^{-1}(u,v)L_{1 j}(u)
=\tilde Z_{21}^{-1}(u,v)L_{1 j}(u)\tilde Z_{21}(u,v)
 L_{2 j+1}(v)\tilde Z_{21}^{-1}(u,v)
\ll{zlzl1u}\end{equation}
  at   nearest neighbour
points and
\begin{equation}
 L_{2 k}(v)\tilde Z_{21}^{-1}(u,v)L_{1 j}(u)
=\tilde Z_{21}^{-1}(u.v)L_{1 j}(u)\tilde Z_{21}(u,v)
 L_{2 k}(v)\tilde Z_{21}^{-1}(u,v)
\ll{zlzl2u}
\end{equation}
with $k>j+1$ 
  answering for the nonnearest neighbours.
Note that along with  the usual quantum $ R_{12}(u-v)$-matrix like
(\re{R-mat})  
additional  $ \  \tilde Z_{12} , \ Z_{12}$ matrices 
 appear, which can be (in-)dependent of the spectral parameters
and 
satisfy  a system of Yang-Baxter type relations
 \cite{kunhla}. 
Due to appearance of $Z$ matrices however one faces  initial 
difficulty in trace
factorisation unlike the  ultralocal models.
 Nevertheless, in most cases
 one can  bypass this problem  by
introducing a $K(u)$ matrix and    defining 

$t(u)=tr (K(u)T(u))$ as commuting matrices  \cite{skly-r,kunhla} for
establishing the quantum integrability for nonultralocal models.
Though a wellframed theory for such systems is yet to be achieved
one can
 derive 
  the basic equations for a series of nonultralocal models 
in a rather  systematic way from the  general relations
  (\ref{bqybel}-\ref{zlzl2u})
by 
 paricular explicit choices of  $Z,\tilde Z$ and $R$-matrices
\cite{kunhla,kunalush}. The models which can be covered through 
this scheme are 
\\ 
 1. {\it Nonabelian Toda chain }\cite{natoda}

$\tilde Z=1, Z=I+i\hbar (e_{22} \otimes e_{12}) \otimes \pi.$
\\ 2. {\it Current algebra in WZWN model} \cite{wzwn}
 
 $\tilde Z=1$ and $Z_{12}=R_{q12}^-,$ where $R_{q}^\pm $ is the $\la \rw \pm
\infty $ limit  of the trigonometric $R(\la)$-matrix
\\
 3. {\it Coulomb gas picture of  CFT} \cite{babelon}
 
  $\tilde Z=1$ and
 $Z_{12}=
q^{-\sum_i H_i
\otimes H_i}
$.
\\
 4. {\it Nonultralocal quantum mapping }\cite{Nijhof}

 $\tilde Z=1$ and $Z_{12}(u_2)
 = {\bf 1}+ \frac { h }{u_2}\sum_\alpha^{N-1}e_{N \alpha
}\otimes e_{\alpha N}~~~.$
 \\5. {\it Integrable model on moduli space }\cite{alex}

  $\tilde Z=Z_{12}=R^+_q.$
 \\
6. {\it Supersymmetric models } 

 $Z=\tilde Z=\sum \eta_{\al \bet} g_{\al \bet}$, where $ \eta_{\al \bet}= 
e_{\alpha \alpha}\otimes e_{\beta \beta}$ and $ g=
 (-1)^{\hat \alpha \hat
 \beta}$ with  supersymmetric grading  $\hat \alpha.$
\\
7. {\it Anyonic type SUSY model}

$Z=\tilde Z=  
 \sum \eta_{\al \bet} \tilde g_{\al \bet}$, with  
   $\tilde g_{\al \bet}=e^{i \theta 
\hat \alpha \hat
 \beta}$.
\\ 
 8. {\it Quantum mKdV model  }\cite{kmpl95}
 
 $\tilde Z=1, ~
Z_{12}=  Z_{21}= q^{-\frac {1}{2} \sigma^3\otimes \sigma^3
},$
and  the  trigonometric $R(u)$ matrix.
\\ 9. {\it Kundu-Eckhaus equation} \cite {kun84}

Classically integrable  NLS equation with 5th power nonlinearity  
\be i\psi_1+\psi_{xx}+ \kappa (\psi^\dagger \psi) \psi
+ \theta^2   (\psi^\dagger \psi)^2 \psi +2i \theta
 (\psi^\dagger \psi)_x \psi =0, \ee
as a  quantum model involves   anyonic type fields:
 $ ~\psi_n \psi_m= e^{i \theta} \psi_m\psi_n,~~ n>m;  
~~[ \psi_m, \psi_n^\dagger]=1.$ The choice
 $\tilde Z=1,
Z= diag (e^{i\theta},1,1,e^{i\theta})$ and the rational $R$ matrix
  constructs the braided QYBE,
 The trace factorisation problem
 has not been solved.

Other   models of nonultralocal class 
are the wellknown {\it Calogero-Sutherland} (CS) and 
{\it Haldae-Shastry} (HS) models with interesting long-range interactions.
 Spin extension
  of the  CS 
 model may be given by the Hamiltonian\c{cs,hal-pas} 
\be
H_{cs}= \sum_{j=1}^{N} p^2_j+ 2 \sum_{1\geq j<k \geq N} (a^2-a
P_{jk})V(x_j-x_k)\ll{hscs}\ee
with 
$~[x_j,p_k]= i\de_{jk}~$, where the potential 
 $~V(x_j-x_k)= {1 \ov (x_j-x_k)^2}~$ for nonperiodic and 
$~ {1 \ov \sin ^2 (x_j-x_k)}~$
for the periodic model. 
 $~P_{jk}$ is  the permutation operator  responsible for exchanging the
spin states of the $j$-th and the $k$-th particles. In the absence of the
operator $P_{ij}$, (\re{hscs}) turns into the  original CS model without
 spin.

The spin CS model exhibits many fascinating features, namely its conserved
quantities including the Hamiltonian exhibit Yangian symmetry, the
eigenvalue problem
  can be  solved  exactly using 
Dunkl operators,  
 the ground state is a solution of the Knizhnik-Zamolodchikov
equation, the system can be viewed  as the free anyonic  gas related to
the notion of fractional statistics etc. \c{hal-pas}. Though 
the satisfactory formulation of 
 quantum 
integrability  of the  model by
braided QYBE has not yet been achieved, this was done   through an
alternative procedure using  operators $L$ and $M$ and showing 
$~~ [H_{cs}, L]=[L,M]
~~$
\c{wad-shas}. 

Remarkably, at $a \rw {\infty}$ the Hamiltonian of the CS model (\re{hscs})
for the periodic case reduces to the HS { model}
\c{HS}
\be
H_{hs}=  \sum_{j<k } 
{P_{jk} \ov \sin^2(x_j-x_k)}.\ll{hhs}\ee
This discretized  long-range interacting spin chain like model seems to be
less well understood and its Lax operator description difficult to find.

\section {Interrelation of QIS with other fields}
As I have  mentioned in the introduction   
the quantum integrable system is 
intimately connected with various other branches of physics and mathematics.
Therefore the knowledge and techniques of QIS is often helpful in
understanding and solving other problems.
Here we briefly touch upon some of these relations just to demonstrate 
the wide range of applicability of the theory of integrable systems.
\subsection { Relation with statistical systems} 
The $(1+1)$ dimensional quantum systems are linked  with $2$
dimensional classical statistical systems and the notion of integrability is 
equivalent in both these cases. For  integrable statistical systems
 the QYBE (\re {qybel}) and  
 the  YBE (\re {ybe}) becomes the same and leads similarly to the commuting
transfer matrix $\tau(\la)$ for different $\la$.  

Let us  examine a classical  statistical model known as vertex model by
 considering  $2$-dimensional array of $N\times M$
lattice points connected
by the bonds assigned with +ve (-ve) signs or equivalently, with right, up
(left, down) arrows  in a random way.
 The
partition function $Z$ of this system 
may be given starting from  local properties, i.e.  by finding the probability
of occurrence of a particular configuration at a fixed lattice point $i$.
 For 
 two   allowed signs on each bond,
 $4\times 4=16$  possible
arrangements arises at each lattice point.
 Setting the corresponding Boltzmann
weights $w_j=e^{-\epsilon_j\beta}$ as the matrix elements of a
 $4\times4$-matrix, we get the $R^{(i)}_{12}
$-matrix with crucial dependence  
 on  spectral  parameter $\la$.
 The configuration 
 probability for a string of $N$-lattice 
 sites in a row may be given by  the transfer
matrix 
$ \tau(\vec {\alpha},\vec{\beta})
= tr (\prod_i^N R^{(i)})$.
 For calculating   the partition function
involving $M$ such strings, one
has to repeat  the  procedure $M$ times to give 
$Z = tr (\prod^M T )=  tr ( T^M)
$.

The YBE (\re {ybe}) restricts the solution of the $R$-matrix 
 to  integrable models. However  
the $R$ matrix with $16$ different Boltzmann weights, 
 representing in general
a $16$-vertex model is difficult to solve. Therefore 
we  imposed  some extra symmetry and conditions 
 on the $R$-matrix 
 by  requiring
 the  { charge} conserving symmetry $R^{ij}_{kl} \neq 0, ~$, only  when 
$k+l=i+j$,
  along 
 with   a charge or arrow
  reversing  symmetry (see fig. 1 on page 16).

 Using an overall normalisation it leads to  a $6$-vertex model
for which the $R$-matrix is given exactly by (\ref{R-mat}), which in turn
represents the Lax operator of the $XXZ$ spin-$\ha$ chain as constructed
above.
Thus we see immediately the similarity between statistical and the 
quantum systems in their  construction of  $R$-matrices, transfer matrix,
integrability equations etc. This deep 
analogy goes also through all the steps in
solving the eigenvalue problem by Bethe ansatz in both the systems, e.g. 
vertex models in integrable 
statistical systems and the spin chains in QIS \c{baxter,xyz}.
\subsection {interrelation between QIS and CFT}
There exists deep interrelation between these two two-dimensional systems,
first revealed perhaps by Zamolodchikov \cite{zam} by   showing that, 
if CFT is perturbed through relevant perturbation and the system goes away
from criticality it might generate hierarchies of integrable systems. For 
example $c=\ha$ CFT perturbed by the field $\si=\phi_{(1,2)} $ as
$H=H_{\ha}+h \si \int \si(x) d^2x, $ represents in fact the ising model at
$T=T_c$ with nonvanishing magnetic field $h$. Similarly  the 
WZWN model perturbed by the operator $  \phi_{(1,3})$ generates  integrable
restricted sine-Gordon (RSG) model.
Under such perturbations the trace of the tress tensor, unlike pure CFT,
 becomes nonvanishing and generates in principle infinite series of
integrals of motion associated with the integrable systems.

In recent years this relationship has also been explored by 
 streching it in a sense from the opposite direction.
The aim was to describe
   CFT  
through {\it massless} $S$-matrix \c{BLZ}
starting from the theory of integrable systems .
 This alternative approach    
  based on the quantum KdV model
attempts to  capture the integrable structure of  CFT.   
Note that the conformal symmetry of CFT is generated by its energy-momentum
tensor $~ T(u)=-{c \ov 24}+\sum_{-\infty}^{\infty} L_{-n}e^{inu}~,$ with
$L_n$ satisfying the Virasoro algebra. The operators 
$I_{2k-1}= {1 \ov 2 \pi} \int_0^{2 \pi} du T_{2k}(u)~$ with $
 T_{2k}(u),$  depending on various
 powers and derivatives of $ T(u)$ represents an
infinite set of commuting integrals of motion. The idea is to solve their 
simultaneous diagonalisation  problem, much in common to the QISM
 for the
integrable theory. Remarkably, this is equivalent  to  solving the quantum
KdV problem, since  at the classical limit the field  $T(u)=-{c
\ov 6} U(u) $ with $U(u+2 \pi)=U(u)$  reduces the commutators of $T(u)$ to
$ \ \{U(u),U(v)\}=2(U(u)+U(v))\de'(u-v)+\de'''(u-v) ,\ $
which is the well known Poisson structure of the KdV.   

Another practical application of this relationship 
 is  to extract the important information about
the underlying CFT  in the scaling limit  of the
integrable lattice models. Interestingly, from the finite size correction
of the Bethe ansatz solutions, one can determine \c{karow} the CFT
characteristics like the central charge and the  conformal dimensions.
For example 
one may analyse the finite size effect of the Bethe ansatz solutions
of the six-vertex model ( with a seam 
given by  $\kappa$). Considering the coupling parameter 
 $q=e^{i{ \pi \ov \nu+1}},$ one obtains from the Bethe solution 
 at the large
$N$ limit  the expression
\[ E_0=Nf_{\infty}-{1 \ov N}{\pi \ov 6} c + O({1 \ov N^2}) \]
 for the ground state energy
and
\[ E_m -E_0 ={2 \pi \ov N}( \De+ \ti \De) + O({1 \ov N^2}) \qquad
 P_m -P_0 ={2 \pi \ov N}( \De- \ti \De) + O({1 \ov N^2}) \]
 for the excited states. Here 
  $\De,  \ti
\De$  are conformal weights of  unitary minimal
models  and 
$c=1-{6 \kappa^2 \ov \nu(\nu+1)}, \ \nu=2,3, \ldots$
is the central charge of the  corresponding conformal field theory.  

\subsection{Link polynomial using integrable systems}

A link polynomial is an invariants corresponding to a  particular knot
or link and is extremely useful for classifying them. 
Jones polynomial is such an example. There are various ways
to construct such    polynomials. Interestingly the Integrable systems 
provide a systematic highly efficient way of producing such polynomials,
which can distinguish between different knots, where even Jones polynomial
fails.
The main idea is to start with a trigonometric  $R(\la)$-matrix  solution 
of the YBE, which in general is a $N^2 \times N^2$ matrix depending on the
higher representation of the  $SU(2)$ algebra. Then the task is to find 
the corresponding braid group representation by taking the $\la \rw \infty $
limit. Defining now the {\it Markov trace} in a particular way one can
construct a series of link polynomials for different cases of $N=2,3
\ldots$ .
Higher the $N$  richer is the contents of the polynomial. 
For example, using $N=2$ one gets the same polynomial for the Birman's two 
closed braids while $N=3$ by the above method generates two distinct
polynomials for these braids \c{wadati}.

\section {Conclusion
without conclusion}
Basic notions  of the quantum integrable systems are explained focusing 
on various aspects and achievements of this theory. The deep interrelations
of this subject with many other fields of physical and mathematical sciences 
are mentioned. However it is difficult yet to draw any conclusion at this
stage, since we expect to hear many more surprises in this evergrowing
field. The recent Seiberg-Witten theory might be one of them.
The influences of this theory in explaining high $T_c$-superconductivity
\c{hightc},
reaction-diffusion processes \c{reacrit} etc. are being felt.
 We expect also
to have breakthrough of quantum integrability in genuine higher dimensions.
Therefore let us leave this conclusion without concluding  and 
keep this task  for the future.

\hrule

$$                  $$
\par
\epsfxsize=6in
\epsfysize=2.5in
\epsffile{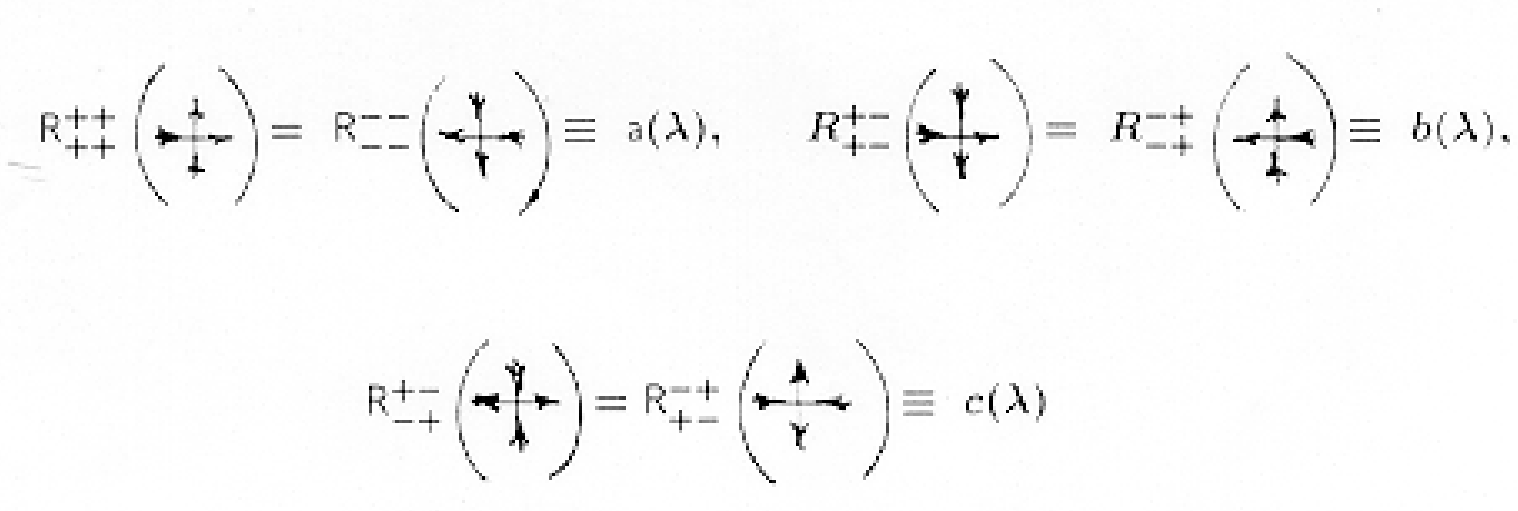}
\par

\vspace*{1cm}

\smallskip

\noindent {\bf figure 1}: {\it Boltzmann weights of the $6$-vortex model constituting the
elements of the $R$-matrix}

\smallskip


\begin{thebibliography}{99}
\bibitem{brussell} 
Barton Russel , {\it Mysticism  \& Logic}, ch. 5
\bibitem{russel} 
R. K. Bullough and P. J. Caudrey, in {\it Solitons} 
 ( Springer-Verlag, 1980) 
\bibitem{fup} E. Fermi, J. R. Pasta and S. M. Ulam,
Collected Works of E. Fermi, Vol 2 (Univ. of Chicago Press, 1965) p. 978
\bibitem{krus65} N. Zabusky and M. D. Kruskal, Phys. Rev. Lett. 15 (1965)
240 
\bibitem{soliton} S. Novikov, V. Manakov, L. Pitaevskii and V.  Zakharov,
Theory of Solitons (Plenum, N.Y., 1984); 
M J Ablowitz, D J Kaup, A C Newell and H Segur, Phys Rev Lett, 31 (1973)
125
\bibitem{lakh93} M. Lakshmanan and R. Sahadeban, Phys. Rep. 224 (1993) 1 
\bibitem{fadrev} L. D. Faddeev, Sov. Sc. Rev. C1 (1980) 107.
\bibitem{qism}
H. B. Thacker, Lect. Notes in Phys. vol. 145 (Springer, 1981), 1:
 Rev. Mod. Phys. 53 (1981) 253:
J H Lowenstein in Les Houches Lect. Notes (ed. J B Zuber et al, 1984)p. 565
; P. Kulish and E. K. Sklyanin,
Lect. Notes in Phys. (ed. J. Hietarinta et al, Springer,Berlin, 1982) vol. 151
p. 61.
\bibitem{majid} S. Majid, J. Math. Phys. 32 (1991) 3246
\bibitem{maillet} L. Freidel and J.M. Maillet, Phys. Lett. 262 B  (1991) 278.
 : Phys. Lett. 263 B  (1991) 403
\bibitem{kunhla}
L. Hlavaty  and Anjan Kundu, { Int J. Mod. Phys.}
  11 (1996) 2143
\bibitem{natoda} V. E. Korepin, J. Sov. Math. 23 (1983) 2429

\bibitem{Nijhof} F.W. Nijhoff, H.W. Capel and V.G. Papageorgiou,
Phys. Rev. A 46 (1992) 2155

\bibitem{babelon} O. Babelon and L. Bonora, Phys. Lett. 253 B (1991) 365:
 O. Babelon, Comm. Math. Phys.  139 (1991) 619:
 L. Bonora and V. Bonservizi, Nucl. Phys. B 390  (1993) 205
\bibitem{alex} A. Yu. Alexeev, {\it Integrability in the Hamiltonian
Chern-Simons theory}, preprint hep-th/9311074 (1993).
\bibitem{construct}
     Anjan Kundu and B. Basumallick   {
     Mod. Phys. Lett. A} { 7} (1992) 61
\bibitem{kun84} Anjan  Kundu J. Math. Phys. 25 (1984) 3433
:
F. Calogero, Inverse Prob. 3 (1987) 229
:
L. Y. Shen, in {\it Symmetries and Singularity Structures} (ed. M.
Lakshmanan, Springer Verlag, NY, 1990) p. 27
\bibitem{kmpl95} 
 Anjan Kundu,  Mod. Phys. Lett.  A 10 (1995) 2955
\bibitem{kunalush} 
 Anjan Kundu in Prob. of QFT (D.V Shirkov et al, JINR publ.,Dubna,1996) p. 
140
\bibitem{skly-r} E. Sklyanin, J. Phys. A 21 (1988) 2375
\bibitem{baxter} R. Baxter,
 {\it Exactly solved models in statistical mechanics}
(Acad. Press, 1981)
\bibitem{xyz} L.A. Takhtajan  and L.D. Faddeev, Russian Math. Surveys 34
(1979) 11
\bibitem{BLZ} V. V. Bazhanov, S. L. Lukyanov and A. B. Zamolodchikov,
Comm. Math. Phys. 177 (1996) 381:
V. A. Fateev and S. Lukyanov, Int. J. Mod. Phys. A7 (1992) 853, 1325
\bibitem{zam}  A. B. Zamolodchikov,
Pisma ZETF  46 (1987) 129
\bibitem{wadati} M. Wadati, T. Deguchi  and Y. Akutsu,  
 { Phys. Rep.}
 {180} (1989) 247
\bibitem{hightc} Haldane,avchinnikov,wilson  
\bibitem{reacrit}
F. C. Alcaraz, M. Droz, M. Henkel, V. Rittenberg, Ann. Phys. 230 (1994) 667
\bibitem{karow} A. Karowski, Nucl. Phys. B 300 [FS 22] (1988) 479
H. J. de Vega and  A. Karowski, Nucl. Phys. B 285 [FS 19] (1987) 619
\bibitem{wzwn}
L. D. Faddeev,  Comm. Math. Phys.  132 (1990) 131
A. Alekseev, L.D. Faddeev, M. Semenov-Tian-Shansky  and
A. Volkov,{\it The unraveling of the quantum group structure in the
WZWN theory},  preprint CERN-TH-5981/91 (1991)
\bibitem{lipatov} L. N. Lipatov, Phys. Lett. B309 (1993) 394
; Phys. Rep. 286 (1997) 131
\bibitem{liuFad}  L. D. Faddeev and O.  Tirkkonen, Nucl. Phys. B453
(1995) 647
\bibitem{hal-pas} D. Bernard, M. Gaudin, F. Haldane, V. Pasquier, J. Phys.
A 21 ? (1993) 5219

\bibitem{wad-shas} K. Hikami and M. Wadati, J. Phys. Soc. Japn 62 (1993) 469
:
B. Sutherland and S, Shastry, Phys. Rev. Lett. 71 (1993) 5
\bibitem{cs} F. Calogero, J. Math. Phys. 12 (1971) 418
:
J. Moser,  Adv. Math. 16 (1975) 197
:
B. Sutherland, Phys. Rev. A5 (1972) 1372
\bibitem{HS} F.D.M. Haldane, Phys. Rev. Lett. 60 (1988) 635
:
B.S. Shastry, Phys. Rev. Lett.   60 (1988) 639
\end{thebibliography}
 \end{document}